# Manipulation of the large Rashba spin splitting in polar two-dimensional transition metal dichalcogenides

Qun-Fang Yao[1], Jia Cai[1], Wen-Yi Tong[1], Shi-Jing Gong[1]*,
Ji-Qing Wang[1], Xiangang Wan[2], Chun-Gang Duan[1,3], J. H. Chu[1,3]

[1]*Key Laboratory of Polar Materials and Devices, Ministry of Education,
East China Normal University, Shanghai 200062, China*

[2]*National Laboratory of Solid State Microstructures and Department of Physics,
Nanjing University, Nanjing, Jiangsu 210093, China*

[3]*Collaborative Innovation Center of Extreme Optics, Shanxi University,
Taiyuan, Shanxi 030006, China*

**ABSTRACT**

Transition metal dichalcogenide (TMD) monolayers MXY (M = Mo, W; X ≠ Y = S, Se, Te) are two-dimensional polar semiconductors. Setting WSeTe monolayer as an example and using density functional theory calculations, we investigate the manipulation of Rashba spin orbit coupling (SOC) in the MXY monolayer. It is found that the intrinsic out-of-plane electric field due to the mirror symmetry breaking induces the large Rashba spin splitting around the Γ point, which, however, can be easily tuned by applying the in-plane biaxial strain. Through a relatively small strain (from -2% to 2%), a large tunability (from around -50% to 50%) of Rashba SOC can be obtained due to the modified orbital overlap, which can in turn modulate the intrinsic electric field. The orbital selective external potential method further confirms the significance of the orbital overlap between W-$d_{z^2}$ and Se-$p_z$ in Rashba SOC. In addition, we also explore the influence of the external electric field on Rashba SOC in the WSeTe monolayer, which is less effective than strain. The large Rashba spin splitting, together with the valley spin splitting in MXY monolayers may make a special contribution to semiconductor spintronics and valleytronics.

# I. INTRODUCTION

Since the successful exfoliation of graphene by Novoselov *et al.* in 2004,[1] growing research attention has been focused on the two-dimensional materials, which consequently accelerates the emergence of other two-dimensional materials, such as boron nitride (BN),[2] silicene,[3] and transition-metal dichalcogenide (TMD) monolayers $MX_2$ (M = Mo, W; X = S, Se, Te),[4,5] *etc*. Because of the intrinsic band gap about 1.1~1.9 eV,[5,6] TMD monolayers are considered to be good candidates for the channel materials in field effect transistor (FET), as well as promising materials for optoelectronics.[7-9] In addition, the inversion symmetry breaking together with the giant spin orbit coupling (SOC) originated from the *d*-orbitals of the metal atoms in TMD monolayers induces the large spin splitting from 150 meV to nearly 500 meV at the corners of the two-dimensional hexagonal Brillouin zone.[10-13] The strong coupling between spin and valley degrees of freedom makes TMD monolayers the ideal valleytronic materials.[14-16]

Different from $MX_2$ monolayers, polar MXY (M = Mo, W; X ≠ Y = S, Se, Te) monolayers can show additional Rashba spin splitting[17] around the $\Gamma$ point, due to the intrinsic out-of-plane electric field induced by the mirror symmetry breaking. According to Cheng *et al.*'s report, Rashba SOC strength in MXY monolayers is around 0.01 eV Å.[10] Rashba SOC was initially investigated in semiconductor heterostructures,[18-24] and wins the growing research interest because of its gate tunability[25] and its great significance in the spin FET,[26] in which the spin precession can be electrically controlled in a precise and predictable way.[27,28] Great efforts have been made to overcome the several fundamental challenges in the spin FET, such as the low spin-injection efficiency, the spin relaxation, and the control of spin precession.[28] Recently, an all-electric and all-semiconductor spin FET has been experimentally realized based on Rashba SOC.[29] The polar two-dimensional MXY monolayers with the intrinsic structure inversion asymmetry will surely enrich the family of Rashba SOC and possibly promote the progress of the spin FET, it is therefore necessary to explore the tunability of Rashba SOC in these materials. Since TMD monolayer has three atomic layers in its unit cell, the in-plane strain will certainly result in the change of bonding angles and lengths, which could dramatically influence the electronic structure.[30-39] For example in $MoS_2$ monolayer, there exists a direct-to-indirect band gap transition under ~2% tensile strain,[30,35-37] and a semiconducting to metal transition under 10~15% tensile strain.[38,39] For MXY monolayer, it is expected that the in-plane strain can effectively manipulate Rashba SOC, which is of great significance for both the fundamental physics and the potential application in the spin FET.

In the present work, we investigate the influence of the biaxial strain on Rashba SOC of MXY monolayers. Since the physics in MXY monolayers is essentially the same,[10] we select WSeTe monolayer as an example to demonstrate the tunability of Rashba SOC. It should be noted that we get much larger Rashba SOC strength for MXY monolayers, compared with Cheng *et al.*'s report.[10] The Rashba SOC strength for the WSeTe monolayer is up to 0.92 eV Å in our present investigations, while it is only 0.014 eV Å in Cheng *et al.*'s report.[10] By using the first-principles density-functional theory (DFT) calculations, we demonstrate the strain dependence of Rashba SOC in WSeTe monolayer. It is found that a relatively small strain (from -2% to 2%) can induce a large tunability (from around -50% to 50%) of Rashba SOC. Our recently developed orbital selective external potential (OSEP) method [40, 41] reveals that the modified orbital overlap between W-$d_{z^2}$ and Se-$p_z$ plays the critical role in manipulating Rashba SOC. We also investigate the influence of the external electric field on Rashba SOC, which is found less effective than the strain.

## II. METHODOLOGY AND MODEL

We perform the first-principles calculations within DFT as implemented in the Vienna Ab-initio Simulation Package (VASP).[42] Since WSeTe monolayer is a polar material, we consider the dipole correction in the calculations, which is introduced by adding an external dipole layer in the vacuum region.[43] In order to eliminate the interaction between adjacent monolayers, a large enough vacuum thickness (~20 Å) along the z axis is adopted. The exchange correlation potential is treated in the local density approximation (LDA). The surface Brillouin zone is sampled with k-points meshes 15×15×1, and the energy cutoff is set to 500 eV for the plane wave expansion of the projector augmented waves (PAWs) in the self-consistent calculations. The convergence of the total energy has been checked by changing the number of sampling k-points, energy cutoff, and the thickness of vacuum space. The structures are relaxed until the Hellmann-Feynman forces on each atom are less than 1 meV/Å. After the structure optimization, we also confirm the stability of the WSeTe monolayer by calculating the phonon dispersion (see supplementary material [44]) with the Phonopy code.[45] In addition, according to Defo *et al.*'s recent report, polar substrate can be used to stabilize the MXY monolayers.[46]

In addition, to reveal the orbital overlap in the Rashba spin splitting bands, we introduce our recently developed OSEP method.[40, 41] This method can introduce a special external potential on the selected orbitals, which has some similarity with the DFT+*U* method. Within the frame of OSEP, the system Hamiltonian is written as $H^{\text{OSEP}} = H^0_{\text{KS}} + |inlm\sigma\rangle\langle inlm\sigma|V_{\text{ext}}$, where $H^0_{\text{KS}}$ is the primary

Kohn–Sham Hamiltonian, and $V_{ext}$ is the applied potential energy. Index *i* denotes the atom site, and *n*, *l*, *m*, *σ* represent the principle, orbital, magnetic and spin quantum number, respectively. Since the strength of overlap between two orbitals is strongly dependent on their energy difference, we can control the orbital overlap by applying an external field to shift the energy levels of the orbitals.

### III. RESULTS AND DISCUSSION

The atomic structure of the WSeTe monolayer is shown in the inset of Fig. 1(a). For the WSeTe monolayer, the mirror symmetry is broken, which leads to a potential gradient normal to the basal plane. We show the planar average of the electrostatic potential energy in Fig. 1(a), in which $z_0$ is the thickness of the unit cell, $z$ is a coordinate variable, and $z/z_0$ refers to the relative position in the unit cell. Direction of the local electric field between W and Se (Te) is indicated by red (blue) arrow, which points from W to Se (Te). The net electric field shown by the black arrow points from Te to Se, which results in the energy difference between the two vacuum levels. Since the Fermi level is set as zero, such energy difference is actually the work function change Δϕ,[47] which is believed to be directly proportional to dipole moment μ, *i.e.*, Δϕ ∝ μ, according to the Helmholtz equation.[48] In addition, we also calculate the charge difference between WSeTe bulk and its monolayer, through which we can observe the change of the charge density induced by the mirror symmetry breaking. A 1×1×2 WSeTe bulk supercell is considered, which can provide a large enough vacuum thickness when we remove the redundant layers to get the monolayer. Since the bulk supercell and the monolayer have the same volume, we can conveniently get the charge difference between them. The charge difference is shown in Fig.1 (b), in which electron depletion in yellow can be observed around W atom and electron accumulation in cyan can be seen around Te and Se atoms. It is clear that more electrons are accumulated around Se than Te, which results in the net electric field pointing from Te to Se atom.

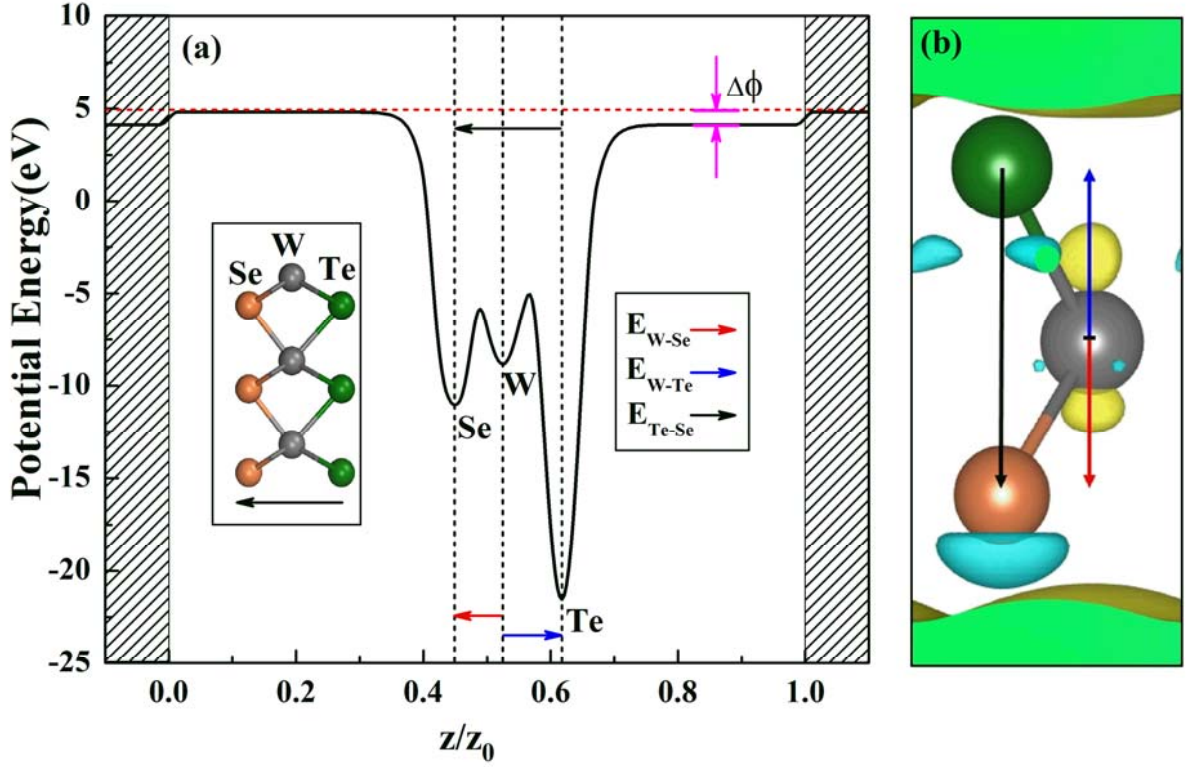

Fig. 1 (Color online) (a) Planar average of the electrostatic potential energy of the WSeTe monolayer, in which $z_0$ is the thickness of the unit cell, z is a coordinate variable, and $z/z_0$ refers to the relative position in the unit cell. The inset in (a) is the side view of the atomic structure of the WSeTe monolayer. (b) The charge density difference between WSeTe bulk and its monolayer, with electron depletion shown in yellow and electron accumulation in cyan. Both in (a) and (b), the green, gray, and orange balls represent Te, W, and Se atoms, respectively. The red (blue) arrow indicates the direction of the local electric field between W and Se (Te) atom, and the black arrow represents the net intrinsic electric field.

To clearly demonstrate the Rashba spin splitting, we show the electronic structures of the WSeTe monolayer with and without SOC in Figs. 2(a) and (b), respectively. The first Brillouin zone and the high symmetry k-points (Γ, K, K', M) are shown in Fig. 2(c), in which the K and K' points are not equivalent due to the threefold symmetry of the WSeTe monolayer. The energy band dispersion is calculated along the selected high symmetry lines Γ−M−K−Γ−K'. According to our calculations, the WSeTe monolayer is an indirect band semiconductor with the valence band maximum (VBM) at the K(K') point and the conduction band minimum (CBM) along K(K')−Γ, which is consistent with the previous report.[10] The irreducible representation at the K point and Γ point is one-dimensional and non-degenerate except for spin, as shown in Fig. 2(a). Due to the spin orbit coupling, the spin degeneracy at the VBM and CBM is removed, and we use the olive and orange curves to highlight the splitting bands. Except the giant valley spin splitting ($\lambda_{kv} \sim 449$ meV) around the K(K′) point, we also get the obvious Rashba splitting around the Γ point. In Fig. 2(c), we plot the distribution of the spin polarization along the ΓK (K′) for the highest valence band, in which the red arrows indicate the in-plane spin polarization, and the blue/yellow contour indicates the out-of plane spin polarization. It can be clearly seen that, from the Γ point to the K(K′) point, the spin polarization turns from in-plane to out-of-plane, and the spin polarization at the K and K′ points has opposite directions. Since WSeTe monolayer has the $C_{3v}$ point symmetry, Rashba spin orbit coupling can be expressed as: $H_{soc} = \alpha_R(k_x\sigma_y - k_y\sigma_x) + \beta_R[(k_x^3 + k_xk_y^2)\sigma_y - (k_x^2k_y + k_y^3)\sigma_x] + \gamma_R(k_x^3 - 3k_xk_y^2)\sigma_z$ ,[49] where the former two items result in the in-plane spin polarization, and the third item results in the out-of-plane spin polarization. The Rashba parameters ($\alpha_R, \beta_R, \gamma_R$) can be analytically solved by the ***k·p*** perturbation theory.[50, 51] However, within the framework of the first-principles calculations, we can get the parameters by numerical fitting. If we are only concerned with the spin splitting around the Γ point, we can fit the Rashba splitting energy by using the polynomial $\alpha_R k + \beta_R k^3$.[52, 53] For WSeTe monolayer, we get the Rashba parameters $\alpha_R = 0.92$ eV Å, $\beta_R = -4.10$ eV Å$^3$.

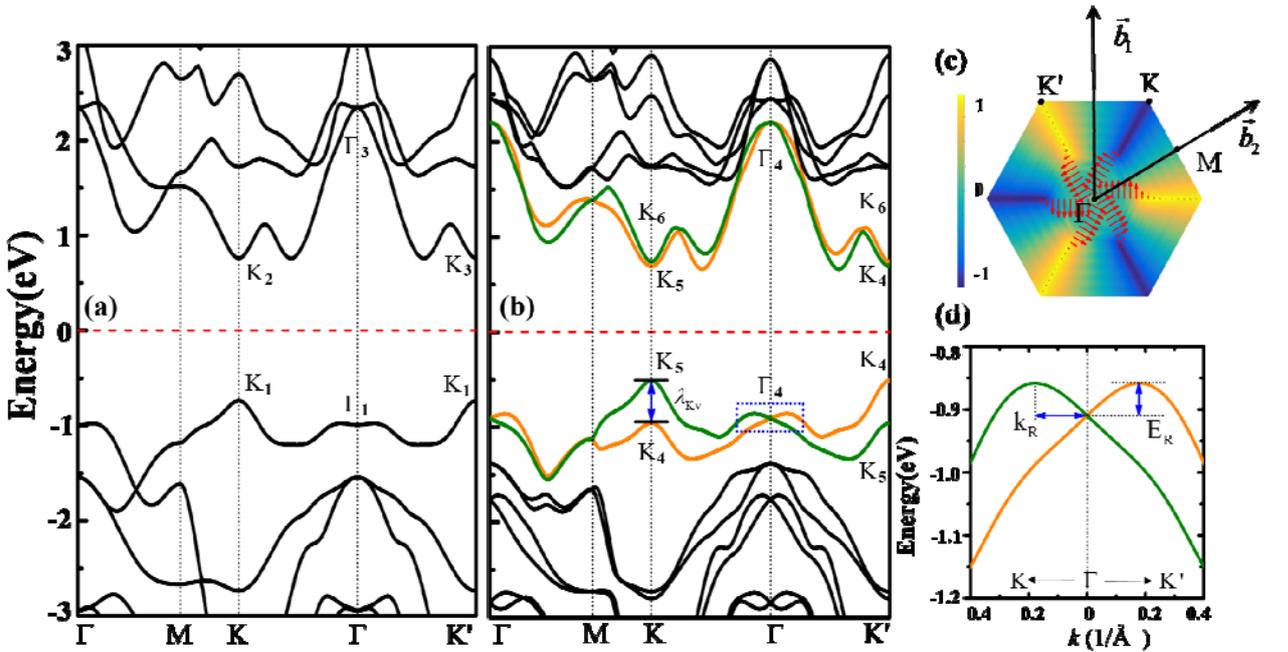

Fig. 2 (Color online) Band structures of the WSeTe monolayer (a) without and (b) with spin orbit coupling. (c) The first Brillouin zone of the WSeTe monolayer with the reciprocal lattice vectors $\vec{b}_1$ and $\vec{b}_2$. The in-plane and out-of-plane spin polarization components on the highest valence band along the lines ΓK (K′) are shown in red arrows and color contour, respectively. (d) Magnified view of the band structure of the highest valence bands around the Γ point.

In Fig. 2(d), we show the magnified Rashba splitting of the WSeTe monolayer, in which Rashba energy $E_R$ and the momentum offset $k_R$ are indicated. We summarize the parameters ($E_R$, $k_R$, and $\alpha_R$) for the WSeTe monolayer and several other materials in Table I. Generally, the large Rashba splitting energy $E_R$ and the momentum offset $k_R$ are desired for stabilizing spin and achieving a significant phase offset for different spin channels. Note that Rashba parameters of the several reference systems listed in Table I are obtained by using the linear Rashba model, within which the energy dispersion for the Rashba splitting bands can be written as $E(k) = \frac{\hbar^2}{2m^*}(|k| \pm k_R)^2 + E_R$, and the Rashba parameter can be obtained from the formula $\alpha_R = \frac{2E_R}{k_R}$. We find that $E_R$ and $k_R$ of the WSeTe monolayer are much larger than those of the traditional semiconductor heterostructure InGaAs/InAlAs,[25] the oxide interface LaAlO$_3$/SrTiO$_3$,[54-56] and the noble metal surface Au(111).[57] Even compared with the newly reported monolayer materials BiTeI monolayer[47] and LaOBiS$_2$,[58] the WSeTe monolayer has the largest Rashba energy and the momentum offset.

Table I. Several selected two-dimensional materials and parameters characterizing the Rashba splitting: Rashba energy $E_R$, the momentum offset $k_R$, and Rashba parameter $\alpha_R$.

| Sample | $E_R$ (meV) | $k_R$ (Å$^{-1}$) | $\alpha_R$ (eV Å) | Reference |
|---|---|---|---|---|
| Au(111) surface | 2.1 | 0.012 | 0.33 | Ref.[57] |
| InGaAs/InAlAs interface | <1.0 | 0.028 | 0.07 | Ref.[25]. |
| LaAlO$_3$/SrTiO$_3$ interface | <5.0 | / | 0.01~0.05 | Ref.[54-56] |
| BiTeI monolayer | 39.8 | 0.043 | 1.86 | Ref.[47] |
| LaOBiS$_2$ | 38.0 | 0.025 | 3.04 | Ref.[58] |
| WSeTe monolayer | 52.0 | 0.170 | 0.92 | This work |

To well understand the Rashba spin splitting in the WSeTe monolayer, we plot the orbital-projected band structures in Fig. 3, with the radius of the circles representing the weight of the orbitals. We notice that the highest valence states around the Γ point are primarily composed of the Se-$p_z$ (blue) and W-$d_{z^2}$ (red) orbitals. The SOC matrix element in the atomic representation can be described by $\xi_l \langle \vec{L} \cdot \vec{s} \rangle_{u,v}$, where $\xi_l$ is the angular momentum resolved atomic SOC strength with $l$ = ($s, p, d$), $\vec{L}$ is the orbital angular momentum operator, $\vec{s}$ is the Pauli spin operator, and $u, v$ indicate the atomic orbitals. As for the WSeTe monolayer, Rashba splitting bands around the Γ point occurs mainly through the SOC matrix elements between the W-$d_{z^2}$ and $d_{xz/yz}$ orbitals, and those between the Se-$p_z$ and -$p_{x/y}$. To directly see the orbital dependence of the Rashba SOC, we artificially switch on or off the partial spin-orbit coupling. Fig. 3(c) shows the spin splitting energy $\Delta E = E^{\uparrow}(k) - E^{\downarrow}(k)$ versus the wavevectors. The red curve in Fig. 3(c) indicates the splitting energy with full SOC, from which we can see that the splitting energy is linearly dependent on the wavevectors around the Γ point. The dashed lines indicate the positions of $\pm k_R$, around which the 'Full SOC' curve has obviously deviated from the linear relation. Switching off SOC of both Se-$p_z$ and W-$d_{z^2}$ orbitals, we notice the spin splitting is drastically suppressed, indicating that these two orbitals play the dominant role in the large Rashba SOC. In addition, we can switch off SOC of Se-$p_z$ (green) and W-$d_{z^2}$ (magenta), respectively. It is found that W-$d_{z^2}$ contributes more to the Rashba spin splitting, and the nonlinear relation between spin splitting and wavevectors is more pronounced in the curve of 'Se-$p_z$: OFF'. This nonlinear relation is also observed in other materials, for example, the narrow gap semiconductor quantum wells,[59] Au(111) surface,[52, 60] bulk BiTeI[61, 62], *etc.*

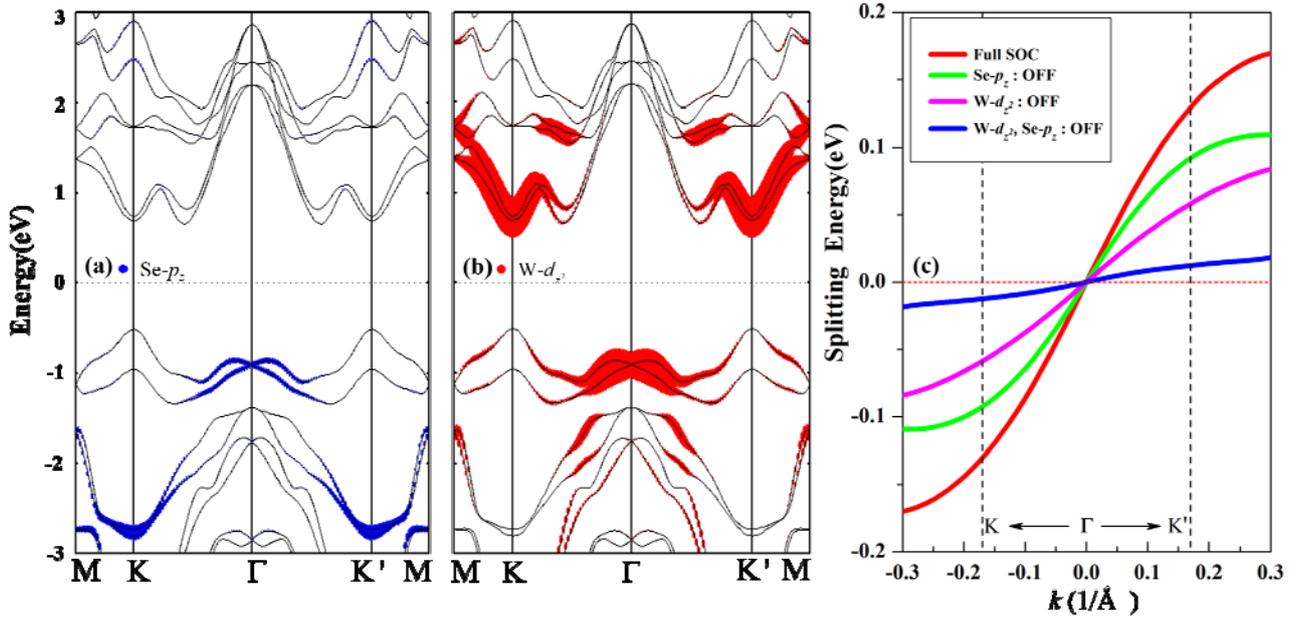

Fig. 3. (Color online) The orbital projected band structures of the WSeTe monolayer, (a) Se-$p_z$ orbital in blue and (b) W-$d_{z^2}$ orbital in red, with the radius of the circles indicating the weight of the orbitals. (c) Spin splitting energy $\Delta E = E^{\uparrow}(k) - E^{\downarrow}(k)$: with full SOC(red), without SOC of Se-$p_z$ (green), without SOC of W-$d_{z^2}$ (magenta), without SOC of both Se-$p_z$ and W-$d_{z^2}$ (blue).

We then apply the in-plane biaxial strain to the WSeTe monolayer. Figs. 4(a)-(e) show the band structures of the WSeTe monolayer under the strain -2%, 0%, 2%, 4%, and 6%, respectively. We notice that, tensile/compressive strain can push up/down the Rashba splitting bands and decrease/increase the corresponding Rashba splitting energy. We believe that the applied strain changes the orbital overlap between atoms, which consequently influences the local electric field and Rashba SOC.[63-66] Checking the work function change Δϕ under different strains, we get the reasonable result that the tensile strain decreases the dipole and the compressive strain increases it. To further confirm the critical role of the orbital overlap, we resort to our recently proposed OSEP method,[40, 41] which has been proved to be a good method to control the overlap between different orbitals, and then reveal the impact of the orbital overlap.[67, 68] For the WSeTe monolayer, we shift the energy level of W-$d_{z^2}$ orbital to tune the overlap between W-$d_{z^2}$ and Se-$p_z$. In Figs. 5(a) and (b), the solid curves show the Rashba splitting bands and splitting energy with $V_{ext}^{d_{z^2}} = \pm 0.6$ and 0 eV, respectively, where the positive (negative) value means shifting up (down) of the W-$d_{z^2}$ orbital. It is clear that with W-$d_{z^2}$ orbital shifting upward, the splitting energy is obviously decreased, and the curves with $V_{ext}^{d_{z^2}}$ = ±0.6 eV fit well with those with ±2% strain. We therefore can conclude that the orbital overlap between W-$d_{z^2}$ and Se-$p_z$ can modulate the Rashba SOC, which acts by modifying the local electric field.

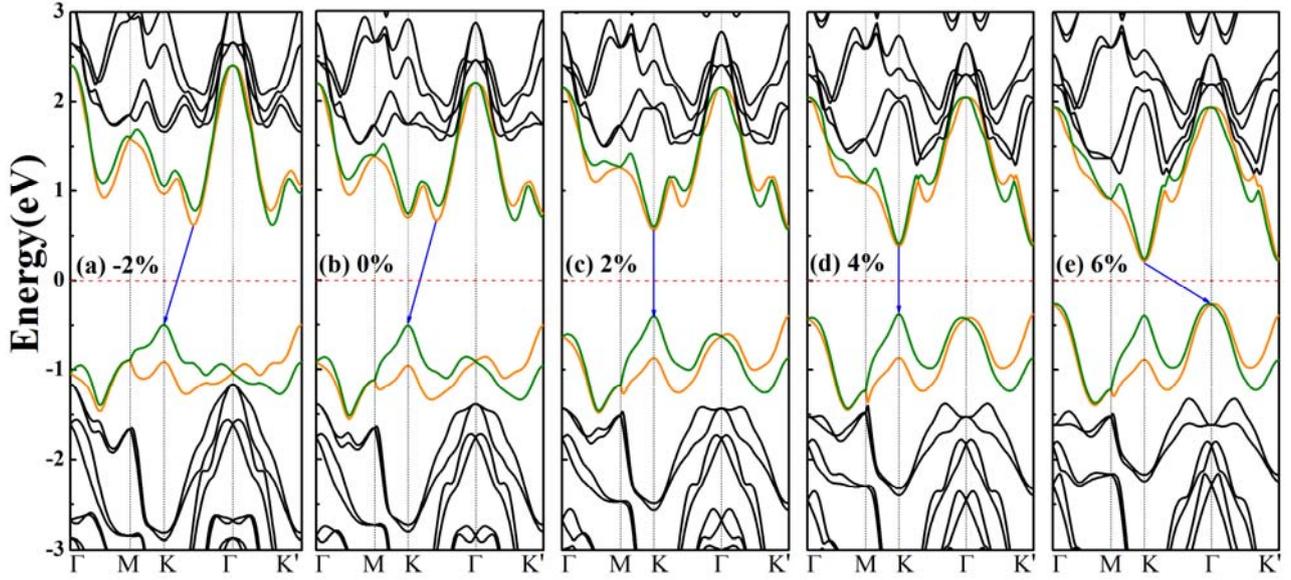

Fig. 4. (Color online) Band structure of the WSeTe monolayer under different biaxial strains -2%, 0%, 2%, 4%, 6%, and the arrows indicate the fundamental band gap.

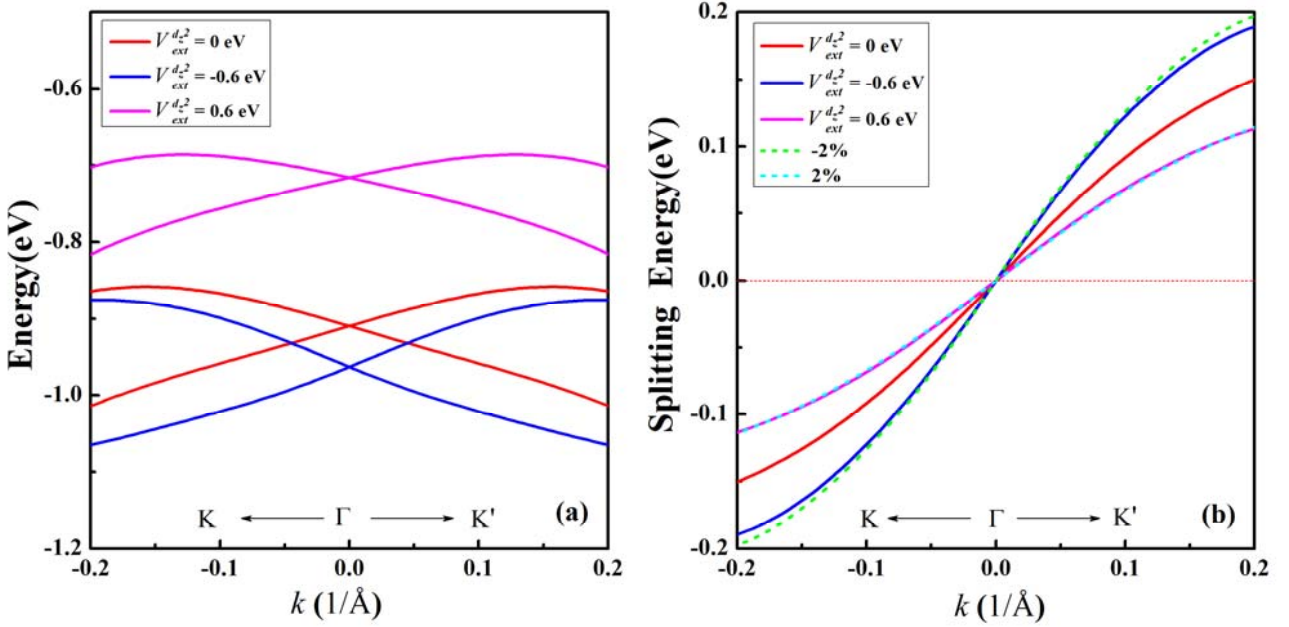

Fig. 5 (a) Rashba spin splitting bands and (b) the corresponding splitting energy, with the externally applied orbital selective potential energy $V_{ext}^{d_{z^2}} = \pm 0.6$ and 0 eV. The dashed curves in (b) indicate the splitting energy with the strain ±2%.

We finally summarize the influence of strain on the band gap and spin orbit coupling parameters in Fig. 6. The band gaps under different strains are shown in Fig. 6(a), in which blue (red) solid triangles indicate indirect (direct) band gap. We find the pristine WSeTe monolayer has the largest band gap, and tensile strain is more effective than compressive strain to reduce the band gap, which is consistent with the previous reports about other TMDs.[34] We also notice that, under the tensile biaxial strain less than 1% the band gap experiences an indirect-to-direct transition, and again, it becomes indirect when the tensile strain increases up to 5%. In Figs. 6(b) - (d), we show Rashba energy $E_R$, the momentum offset $k_R$, and Rashba parameter $\alpha_R$ under different biaxial strains. It can be clearly seen that, these three parameters shown by solid triangles decrease monotonically with the increasing lattice constants, which means that a compressive/tensile strain can enhance/decrease the Rashba SOC strength. From the stripe regions, we can see that with a 2% compressive/tensile strain, Rashba parameter $\alpha_R$ can be increased/decreased by about 50%, which is large enough to effectively tune the spin states. We also look into the change of the higher order Rashba parameter $\beta_R$, and find that its absolute value also increases with the compressive strain and decreases with tensile strain (see supplementary material[44]), which means that if we want a better linear Rashba SOC contribution, we need to balance the linear Rashba SOC strength and the nonlinear Rashba SOC distraction.

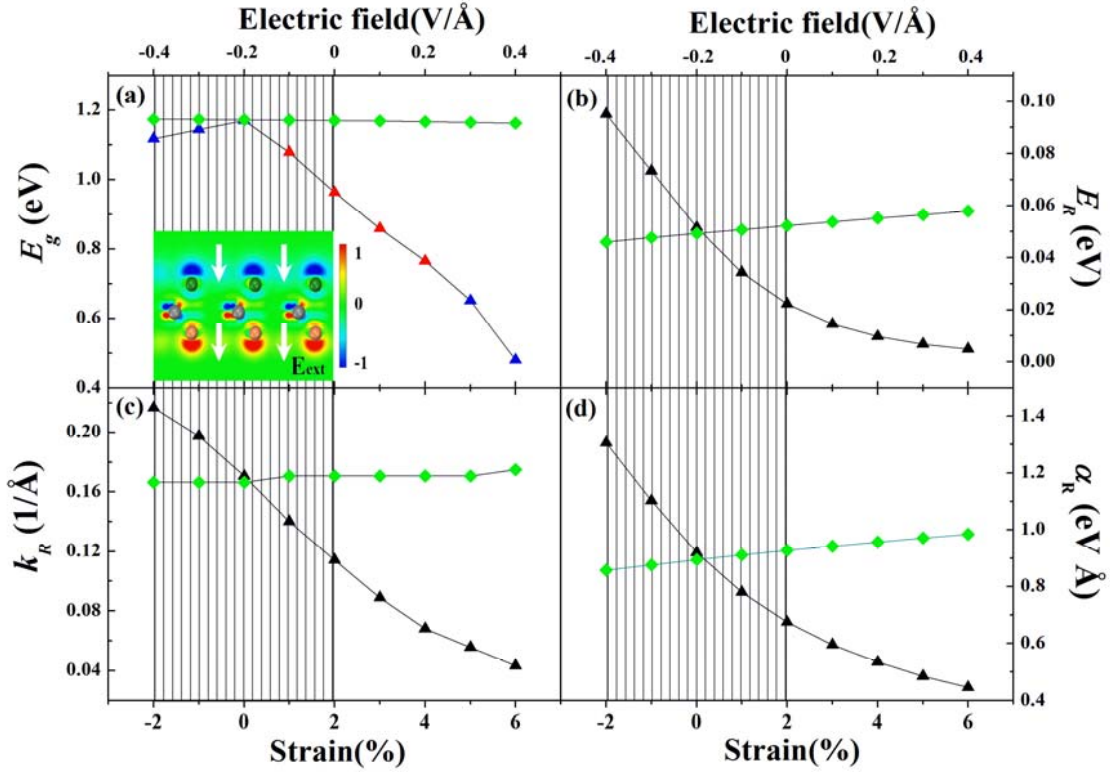

Fig. 6 (a) Energy gap $E_g$ (b) Rashba energy $E_R$ (c) the momentum offset $k_R$ (d) Rashba parameter $\alpha_R$ under different biaxial strains (triangle) and external electric fields (rhombus). In (a), the red triangles indicate the direct band gap, and the blue ones indicate the indirect gap. The inset in (a) is the charge densities induced by the electric field E = 0.4 V/Å, $\Delta\rho = \rho(E_{ext}) - \rho(0)$ in arbitrary units. The vertical stripe regions indicate the strain from -2% to 2%.

Since electric field control of Rashba SOC is of great significance in semiconductor spintronics, we also apply external electric field to the WSeTe monolayer. We consider the external electric field pointing from Te to Se, which is consistent with the local electric field shown in Fig.1. To directly observe the change of charge density induced by the applied electric field, we calculate the charge density difference, *i.e.*, $\Delta\rho = \rho(E_{ext}) - \rho(0)$, and show it in the inset of Fig. 6(a), with $E_{ext}$ = 0.4 V/Å. It is clear that opposite charge densities, which are shown in red and blue, respectively, are induced at the two sides of the WSeTe monolayer. Since the applied electric field has the same direction with the local intrinsic electric field, we hope it can enhance the Rashba SOC. From Fig. 6, we can see that, the applied electric field can only slightly enhance Rashba SOC. This is because the *d*-orbitals of W atom can hardly be influenced by the external electric field due to the screening effect, and meanwhile the atomic SOC of Se atom is not strong. To well understand the electric field influence on Rashba SOC in TMD monolayers, we calculate all the $MX_2$ and MXY monolayers, and find that the electronic states of the anions in the Rashba splitting bands play the critical role. Since the atomic SOC of both S and Se atoms is not strong, among all the TMD monolayers, only $WTe_2$ and $MoTe_2$ monolayers can show obvious Rashba SOC with the assistance of electric field (see the supplementary material[44]). In addition, we also notice that, the band gap of the WSeTe monolayer is insensitive to the applied electric field. It is worthy to mention that, applying electric field to TMD bilayer will induce dramatic change to band gap, even semiconductor–metal transitions could be possible, due to the potential difference between the two layers induced by electric field.[11, 69, 70]

## IV. CONCLUSIONS

Transition metal dichalcogenide monolayers MXY (M = Mo, W; X ≠ Y = S, Se, Te) are two-dimensional polar semiconductors with Rashba spin orbit coupling around the Γ point. Setting the WSeTe monolayer as an example, we explore the tunability of Rashba SOC in MXY monolayer. It is found that the intrinsic out-of-plane electric field in polar WSeTe monolayer induces the large Rashba spin splitting around the Γ point, and the in-plane biaxial strain can effectively tune Rashba SOC by modifying the W-Se bonding interaction, *i.e.*, the orbital overlap, which actually changes the intrinsic electric field. Even through a relatively small compressive/tensile strain (from -2% to 2%), a large tunability of Rashba SOC can be obtained. By using the OSEP method, we demonstrate that the change of the orbital overlap can obviously modify the Rashba SOC. We also explore the influence of the external electric field on Rashba SOC in the WSeTe monolayer, which is found less effective than the strain, because it can hardly influence the charge density distribution of W atom due to the screening effect.

## ACKNOWLEDGMENTS

This work was supported by the 973 Programs No. 2014CB921104, 2013CB922301, NSF of Shanghai (No. 14ZR1412700), the NSF of China (No. 51572085 and 11525417). Computations were performed at the ECNU computing center.

*Corresponding author address: sjgong@ee.ecnu.edu.cn